\documentclass[runningheads]{llncs}

\usepackage[T1]{fontenc}

\usepackage{graphicx}
\usepackage{xcolor}
\definecolor{pink}{rgb}{0.95, 0.8, 0.96}
\definecolor{pinkd}{rgb}{0.95, 0.4, 0.96}
\usepackage[textsize=tiny,backgroundcolor=pink,linecolor=pinkd]{todonotes}
\usepackage{enumitem}
\usepackage{listings}
\usepackage{url}
\usepackage[para]{footmisc}
\usepackage{breakurl}
\usepackage{hyperref}
\usepackage{float}
\usepackage{booktabs}
\usepackage[font=scriptsize]{caption} 
\lstdefinelanguage{Turtle}{
    morekeywords={@prefix, rdf:type, rdfs:domain, rdfs:range, rdfs:seeAlso, @base},
    sensitive=true,
    morestring=[b]",
    morecomment=[l]{\#},
    morestring=[b]",
}

\begin{document}

\title{A Knowledge Graph \\ Informing Soil Carbon Modeling}

\author{Nasim Shirvani-Mahdavi\inst{1}\orcidID{0009-0006-2733-2242} \and Devin Wingfield\inst{1} \and \\ Juan Guajardo Gutierrez\inst{1} \and Mai Tran\inst{1} \and Zhengyuan Zhu\inst{1} \and Zeyu Zhang\inst{1} \and \\Haiqi Zhang\inst{1} \and Abhishek Divakar Goudar\inst{1} \and Chengkai Li\inst{1}\orcidID{0000-0002-1724-8278} \and Virginia Jin\inst{2} \and Timothy Propst\inst{2} \and Dan Roberts\inst{2} \and Catherine Stewart\inst{2} \and \\ Jianzhong Su\inst{1} \and Jennifer Woodward-Greene\inst{2}}

\authorrunning{N. Shirvani-Mahdavi et al.}

\institute{University of Texas at Arlington, Arlington TX 76013, USA \and
United States Department of Agriculture, USA \\
\email{idirlab@uta.edu}}

\maketitle             

\begin{abstract}
Soil organic carbon is crucial for climate change mitigation and agricultural sustainability. However, understanding its dynamics requires integrating complex, heterogeneous data from multiple sources. This paper introduces the \sloppy Soil Organic Carbon Knowledge Graph (SOCKG), a semantic infrastructure designed to transform agricultural research data into a queryable knowledge representation. SOCKG features a robust ontological model of agricultural experimental data, enabling precise mapping of datasets from the Agricultural Collaborative Research Outcomes System. It is semantically aligned with the National Agricultural Library Thesaurus for consistent terminology and improved interoperability. The knowledge graph, constructed in GraphDB and Neo4j, provides advanced querying capabilities and RDF access. A user-friendly dashboard allows easy exploration of the knowledge graph and ontology.  SOCKG supports advanced analyses, such as comparing soil organic carbon changes across fields and treatments, advancing soil carbon research, and enabling more effective agricultural strategies to mitigate climate change.

\keywords{Knowledge Graph Construction, Ontology Engineering, Soil Organic Carbon, Carbon Sequestration}
\end{abstract}
\section{Introduction}
Soil is the largest terrestrial carbon sink and contains more carbon than vegetation and atmosphere combined, positioning it as a key player in global efforts to reduce greenhouse gas (GHG) emissions~\cite{lal2004soil,rodrigues2023soil}. 
Soil organic carbon (SOC) is key to soil health, influencing nutrient cycling, water retention, and microbial activity~\cite{kibblewhite2008soil}. 
By enabling the understanding, measurement, and management of SOC dynamics, soil carbon modeling provides the foundation for strategies to sequester carbon, mitigate climate change, and promote sustainable land management~\cite{luo2011modeling}. Modeling SOC dynamics helps target agricultural practices that improve soil fertility, enhance productivity, and increase resilience to climate extremes such as droughts and floods~\cite{ahmed2023climate}. By advancing environmental sustainability, soil carbon modeling supports global food security while reducing the risk of land degradation~\cite{gomiero2016soil}. 

\vspace{-0.5 mm}
Carbon farming practices, such as reduced tillage, crop rotation diversification, cover cropping, and the application of organic amendments, are essential for enhancing SOC levels~\cite{kane2015carbon,paustian2016climate}. By sequestering carbon in soils, land managers can earn carbon credits, providing financial incentives for adopting sustainable practices~\cite{barbato2023farmer}. 
Effective soil carbon modeling sustains the soil credit markets, ensuring accurate accounting of SOC changes and building trust among stakeholders.

Despite the potential benefits, the complexity of soil carbon modeling presents significant challenges, including complex and dynamic soil processes influenced by biological, chemical, and climatic factors, and uncertainties in measurements. Several notable efforts have been made to develop tools for soil carbon modeling and management. The International Soil Carbon Network (ISCN) has created a global database of soil carbon measurements, facilitating data sharing and analysis across diverse ecosystems~\cite{nave2016international}. Similarly, the Soil Survey Geographic Database (SSURGO)\footnote{\scriptsize\url{https://websoilsurvey.nrcs.usda.gov/}} provides detailed spatial data on soil properties across the United States, supporting site-specific analyses and modeling efforts. The Rapid Carbon Assessment (RaCA) dataset\footnote{\scriptsize
\href{https://www.nrcs.usda.gov/resources/data-and-reports/rapid-carbon-assessment-raca}
{\texttt{https://www.nrcs.usda.gov/resources/%
\linebreak[1]data-and-reports/%
\linebreak[1]rapid-carbon-assessment-raca}}}
 offers a robust inventory of soil organic and inorganic carbon stocks across over 6,000 U.S. locations, providing critical data for carbon modeling and management.

This paper presents the development of the Soil Organic Carbon Knowledge Graph (SOCKG), a semantic infrastructure designed to support robust and user-friendly soil carbon modeling. The primary objective of this knowledge graph is to aid the investigation of factors influencing SOC dynamics and other critical soil measurements. By facilitating carbon sequestration in agricultural soils at large, SOCKG offers support for mitigating climate change and enhancing farm productivity and sustainability. In summary, this work contributes the following:
\begin{itemize}[noitemsep,wide,topsep=0pt]
    \item \textbf{Ontology development:} An ontological model semantically aligned with the National Agricultural Library Thesaurus (NALT)\footnote{\scriptsize\url{https://agclass.nal.usda.gov/}}, ensuring dataset interoperability and consistent terminology for soil carbon research.
    \item \textbf{Knowledge graph construction:} Integration and mapping of SOC data from Agricultural Collaborative Research Outcomes System (AgCROS)\footnote{\scriptsize\url{https://doi.org/10.15482/USDA.ADC/1529828}}, which contains detailed information collected from across 58 experimental fields over a span of 45 years, including soil properties (e.g., bulk density, pH) and management practices (e.g., crop rotation). The knowledge graph organizes this data to support the analysis of SOC dynamics and their relationship with soil properties and management practices. 
    \item \textbf{User-friendly tools:} Development of tools and interfaces to improve SOCKG accessibility and usability, including Neo4j and GraphDB hosting of the knowledge graph for advanced querying and visualization, an interactive dashboard and a data cube with  simplified analytical capabilities for diverse stakeholders, and ontology documentation generated using LLM. 
    \item \textbf{Application insights:} Demonstrations of SOCKG's capabilities in supporting SOC-related use cases.
    \item \textbf{Open-source resources:} We have open-sourced several resources at \sloppy \href{https://idir.uta.edu/sockg/}{https://idir.uta.edu/sockg/}. These resources include an ontology in the Turtle\footnote{\scriptsize\url{https://www.w3.org/TR/rdf12-turtle/}} format, the data graph in formats such as Turtle, N-triples\footnote{\scriptsize\url{https://www.w3.org/TR/rdf12-concepts/}} and Neo4j dump, the source code for preprocessing the source AgCROS data file and populating the knowledge graph with the prepocessed data, and the source code of the dashboard. Additionally, we provide detailed documentation on design and implementation of SOCKG and its various tools. 
\end{itemize}

The paper is structured as follows. Section~\ref{sec:background} reviews background concepts in soil carbon modeling and data. 
Section~\ref{sec:KG-construction} details the development and population of SOCKG, as well as its ontology and data integration. 
Section~\ref{sec:applications} describes the developed query endpoints and dashboard. 
Section~\ref{sec:usecases} presents SOCKG's use cases. 
Section~\ref{sec:conclusion} discusses conclusions and future directions for extending SOCKG’s capabilities.

\vspace{-0.3 cm}
\section{Background}
\label{sec:background}
\vspace{-0.2 cm}
\subsection{Key Concepts in Soil Carbon Modeling} 

\noindent \textbf{Carbon cycle} 
is a fundamental Earth system process that governs the movement of carbon among the atmosphere, oceans, terrestrial ecosystems, and geological reservoirs~\cite{carbon-cycle,bralower2016overview}. In this cycle, carbon transitions between organic and inorganic forms, playing an important role in maintaining the planet's climate balance~\cite{isson2020evolution}. Terrestrial ecosystems, particularly soils, act as significant carbon sinks~\cite{rodrigues2023soil}. This sequestration process helps regulate levels of atmospheric carbon dioxide (CO$_2$), a major GHG driving global climate change~\cite{rodrigues2023soil}. Human activities such as deforestation, fossil fuel combustion, and unsustainable agricultural practices disrupt this balance, releasing large amounts of carbon into the atmosphere. 
Soils contribute to the carbon cycle primarily through the accumulation of SOC, derived from decomposed plant and microbial matter~\cite{keenan2018terrestrial}. Promoting soil carbon sequestration has emerged as a strategy to mitigate climate change~\cite{fawzy2020strategies}.  

\vspace{1mm} 
\noindent \textbf{Voluntary carbon markets} provide a mechanism for organizations and individuals to offset their unavoidable GHG emissions by purchasing carbon credits or earn revenue by selling the credits~\cite{dawes2023voluntary}. Each credit typically represents the removal or avoidance of one metric ton of GHG~\cite{dawes2023voluntary}. These markets provide farmers and land managers with financial incentives, encouraging the adoption of sustainable practices that improve soil carbon levels~\cite{buck2022soil}. 
By verifying and monetizing the carbon stored in soils, these markets reward sustainable farming practices while fostering broader participation in climate action. The credibility of these markets depend on robust methodologies to quantify, report, and verify SOC changes~\cite{zhang2024enhancing}.
However, predicting changes in soil carbon remains a significant challenge due to the inherent heterogeneity of soils and their complex interactions with environmental factors and land use~\cite{zhang2024enhancing}. Addressing these challenges requires comprehensive and integrated soil carbon modeling approaches to better quantify and manage soil carbon dynamics effectively. 

\vspace{1mm} 
\noindent \textbf{Agricultural practices for soil carbon sequestration} 
focus on minimizing carbon losses and increasing organic carbon inputs. 
Reduced tillage and no-till farming preserve soil structure and limit carbon loss~\cite{mehra2018review,fazli2023cultivating}. 
Cover cropping with legumes or grasses protects soil from erosion and adds organic matter, while crop rotation improves soil biodiversity and nutrient cycling~\cite{zhang2024enhancing}.
Organic amendments, such as compost or manure, directly increase SOC levels and microbial activity~\cite{thangarajan2013role}. 
These practices not only enhance carbon sequestration but also improve soil fertility, productivity, and resilience to climate change. 

\vspace{-0.2 cm}
\subsection{Soil Organic Carbon Experimentation Data}\label{sec:data}
\begin{figure}[h!]
    \centering
    \begin{minipage}{0.53\textwidth}
        \centering
        \vspace{-0.4 cm }\includegraphics[width=\linewidth]{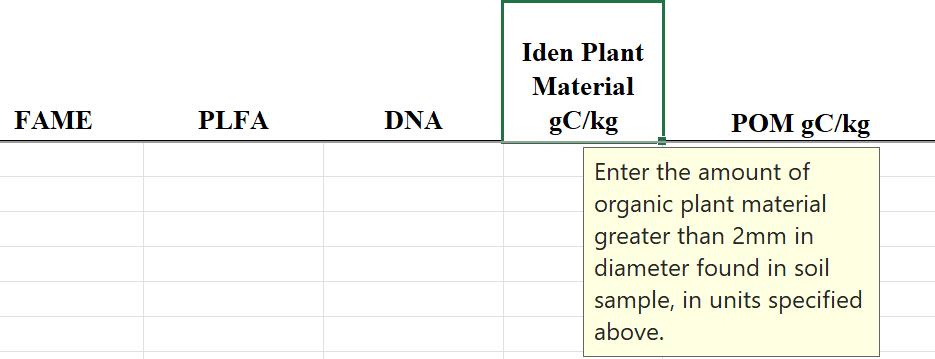}
        \vspace{-1mm}
        \caption{Examples of ambiguous terms and acronyms.}
        \label{fig:ambiguous}
        \vspace{-0.7cm}
    \end{minipage}
    \hfill
    \begin{minipage}{0.43\textwidth}
        \centering
        \vspace{-0.4 cm }\includegraphics[width=0.74\linewidth]{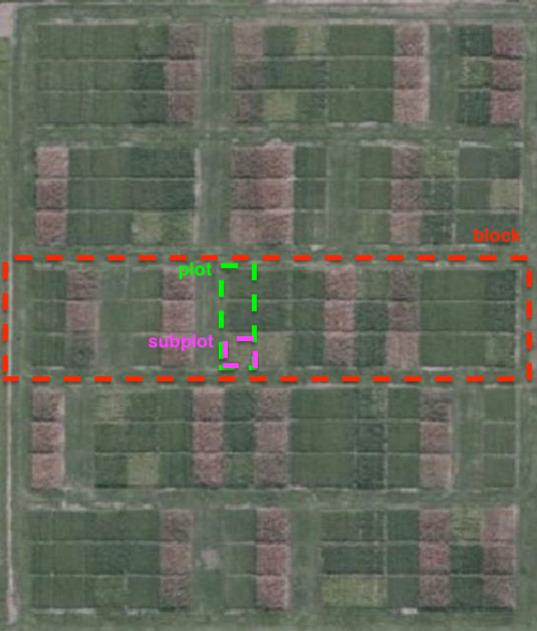}
        \vspace{-0.2cm}
        \caption{An experimental field in NE.}
        \label{fig:exp-site}
        \vspace{-0.7cm}
    \end{minipage}
\end{figure}

The data for this study is derived from the Agricultural Collaborative Research Outcomes System (AgCROS), a repository available by the United States Department of Agriculture (USDA). 
The AgCROS dataset is provided in an Excel spreadsheet, which contains detailed information about various experimental fields. The spreadsheet contains multiple sheets, with most columns associated with contextual notes (an example contextual note in Figure~\ref{fig:ambiguous}, with yellow background) 
which offer additional details, such as measurement scale and range.  The dataset is structured around several key elements, as follows. 
\vspace{-0.35cm}

\vspace{1mm} 
\noindent \textbf{Experimental unit (subplot)}\hspace{0.3cm} It is the basic building block of data collection, organized in a hierarchical structure. Each experimental field is divided into blocks, which are further divided into plots, and then into subplots. The subplots serve as the experimental units where specific treatments are applied and observations are recorded. For example, Figure~\ref{fig:exp-site} illustrates an experimental field in Mead, NE, showing how it is subdivided into these units.

\vspace{1mm}
\noindent \textbf{Treatment}\hspace{0.3cm} It refers to the agricultural practices applied to the experimental units. These practices, designed to assess their effects on soil health and crop performance, include variations in nitrogen levels and crop rotations, accounting for both the crops used and the duration of the rotation cycles.
For example, a treatment with the rotation ``continuous corn'' indicates that corn is the only crop grown in that treatment every year. In contrast, a rotation labeled ``corn/soybean (2-yr)'' signifies that corn is grown in the first year, followed by soybeans in the second year, and this pattern alternates annually, repeating in subsequent years. Nitrogen levels are categorized into three groups: 0, low, and high. However, the specific values for low and high nitrogen levels vary depending on the crop involved in the treatment. For example, for ``corn'' or ``sorghum,'' low nitrogen is defined as 33 kg N/ha, and high nitrogen as 67 kg N/ha. For ``soybean'' or ``oat+clover,'' the corresponding values are 90 kg N/ha and 180 kg N/ha.

\vspace{1mm}
\noindent \textbf{Soil sample}\hspace{0.3cm} Soil samples were collected from various depths in each experimental unit, with each sample defined by its upper and lower depth boundaries. From a given soil sample, one, two or three types of measurements---physical, chemical, and biological---may be taken. However, the AgCROS dataset does not provide information to reflect whether these measurements were taken from the same sample or not. Therefore, a single soil sample in the real world may exist in SOCKG as one, two, or three samples, as follows. A \emph{soil physical sample} is associated with measurements of structural properties of soil such as bulk density. A \emph{soil chemical sample} includes measurements of attributes critical for understanding soil's chemical balance, such as pH and organic matter content. A \emph{soil biological sample} captures information on microbial activity and other biological indicators that reflect soil health and its capacity for carbon sequestration. 

The AgCROS dataset encompasses data collected over a span of up to 45 years from 4,220 experimental units distributed across 58 fields, 33 cities, and 20 states. It includes 37,214 soil physical samples, 77,167 soil chemical samples, and 19,572 soil biological samples collected from these experimental units. 
\vspace{-0.2 cm}
\section{Knowledge Graph Construction}
\label{sec:KG-construction}
This section outlines the creation of SOCKG, starting with data modeling and ontology development. We then describe how SOCKG was populated, integrated with NALT, and enriched with resolvable Uniform Resource Identifiers URIs\footnote{\scriptsize\url{https://www.rfc-editor.org/rfc/rfc3986}}, along with key statistics to highlight its scale and structure. 

\vspace{-0.2 cm}
\subsection{Data Modeling and Ontology Development}\label{sec:ontology}

\begin{figure*}
    \centering
    \vspace{-0.7 cm}\includegraphics[width=0.85\linewidth]{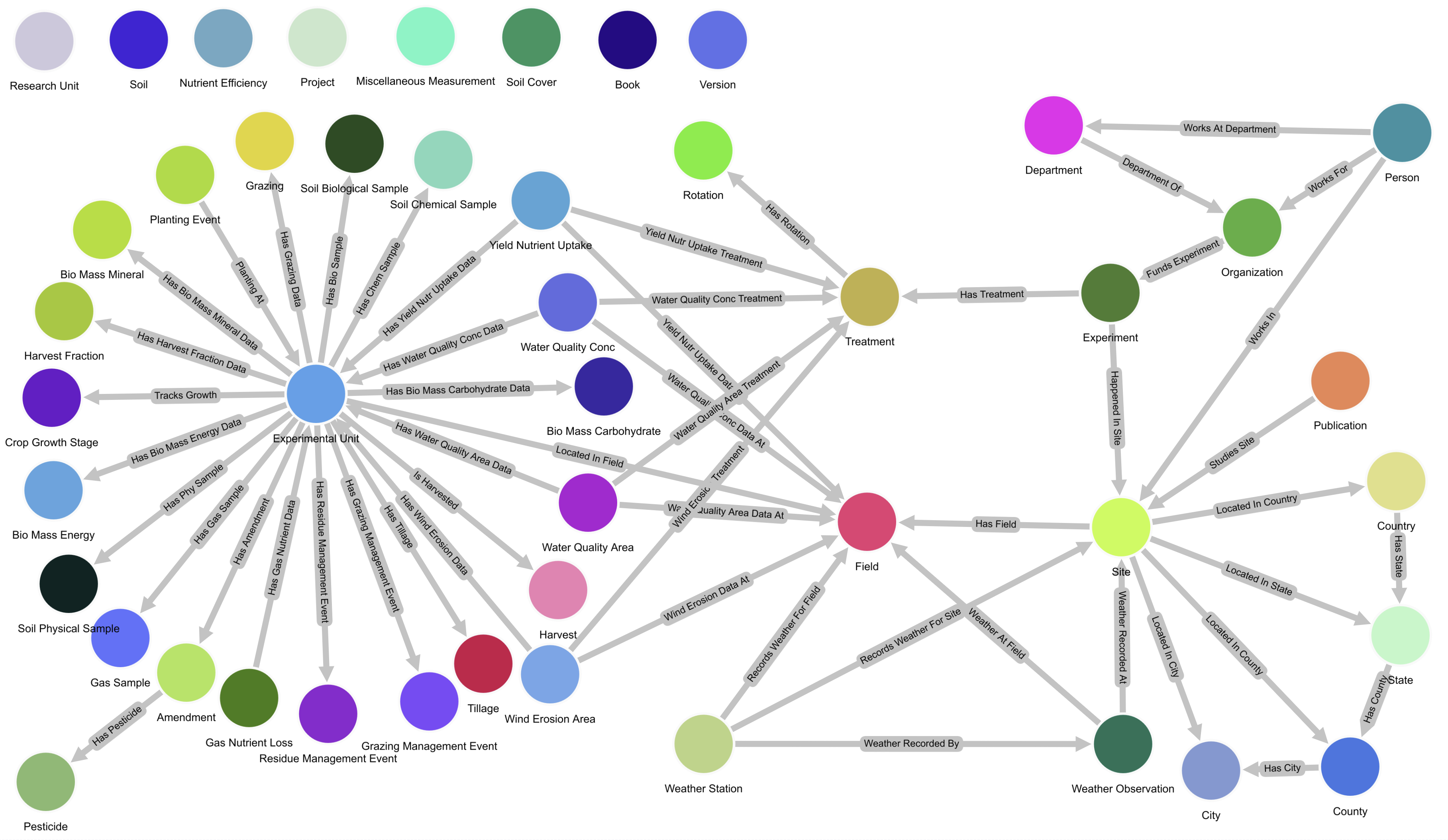}
    \vspace{-2mm}
    \caption{{\scriptsize The SOCKG ontology, depicting classes (nodes) and object properties (edges), excluding data properties. Some nodes (e.g., ResearchUnit) are disconnected due to missing semantic relationships, while synthetic classes (e.g., Version, for tracking dataset history) are inherently unconnected.}}
    \vspace{-0.5 cm}
    \label{fig:dashboard-onto}
\end{figure*}

One of the first and most crucial steps in this work was designing the underlying structure of the knowledge graph, specifically the ontology. The ontology's development was informed by a comprehensive analysis of the dataset and carried out using the ontology editor Protégé~\cite{Protégé}. We used fundamental Web Ontology Language (OWL)\footnote{\scriptsize\url{https://www.w3.org/TR/owl2-overview/}} constructs to define \emph{classes}, \emph{object properties}, and \emph{data properties}.  
Furthermore, the ontology incorporated a specific RDFS relationship, \texttt{rdfs:seeAlso}, to support semantic mapping, as explained in Section~\ref{sec:constExtra}. Figure~\ref{fig:dashboard-onto} provides an overview of the SOCKG ontology, excluding data properties, with nodes in the depicted graph representing classes and edges representing object properties. 
The SOCKG ontology contains 46 classes, 64 object properties, and 590 data properties. Table~\ref{tab:stat} provides more detailed statistics, including the number of data instances for these concepts.

The initial step was to identify the primary concepts represented in the data, which are incorporated into the ontology as classes. This process involved examining the dataset's structure, such as column names and their contextual notes in the AgCROS spreadsheet. For example, concepts such as ``Site,'' ``Field,'' ``Experimental Unit,'' and ``Treatment'' emerged as foundational classes.

\begin{table}
\centering
\vspace{-0.7 cm}
  \caption{Statistics of SOCKG}
  \setlength{\tabcolsep}{10pt}
  \label{tab:stat}
  \scalebox{0.78}{\begin{tabular}{l|c|c|c}
    \toprule
     & \textbf{\# of types in ontology} & \textbf{\# of types in data graph} & \textbf{\# of instances}\\
    \midrule
    \textbf{class} & 46 & 46 & 509,745\\
    \textbf{object property} & 64 & 53 & 738,114\\
    \textbf{data property} & 590 & 489 & 9,934,733\\
  \bottomrule
\end{tabular}}
\vspace{-0.5cm}
\end{table}

Once the main classes were established, data properties were derived by examining attributes that describe each class. Identifiers (e.g., ``Site ID''), numerical metrics (e.g., ``bulk density''  and ``electrical conductivity''), and descriptive attributes were systematically analyzed and assigned to their respective classes. Each data property has a \emph{domain} and a \emph{range}. The domain refers to the class that owns the data property. For example, \sloppy \texttt{bulkDensity\_g\_per\_cm\_cubed} is a physical attribute of soil, hence its domain is \texttt{SoilPhysicalSample}. Conversely, the range defines the type of the data property. Each data property's range was determined by evaluating dataset values and formats. Text attributes (e.g.,  \texttt{crop}) were assigned xsd:string, while numerical values (e.g.,  \texttt{bulkDensity\_g\_per\_cm\_cubed},  \texttt{elevation\_m}) were modeled as xsd:float or xsd:int, as appropriate.

Throughout the ontology development process, attention was given to accurately representing all aspects of the dataset, including attributes that lacked current data. 
There are cases where, while a column header defines a data property, no corresponding data records exist for the column. 
Hence, the data graph contains 489 data properties, compared to 590 data properties in the ontology, as shown in  Table~\ref{tab:stat}.  
In these cases, analysis of the contextual notes found within the dataset and consultation with domain scientists played a critical role in making educated assumptions about the data property's probable range. This ensured that the ontology could accommodate potential future data. 

In addition to data properties, relationships between classes were modeled using object properties. The creation of object properties required an in-depth understanding of the dataset's inherent relationships, often necessitating collaboration with domain experts. For example, relationships between monitoring classes such as experimental units and the phenomena they observe (e.g., soil chemical attributes), were explicitly modeled to reflect their interdependence. One such example is the object property \texttt{hasChemSample}, which has a domain of \texttt{ExperimentalUnit} and a range of \texttt{SoilChemicalSample}, reflecting that an experimental unit observes a soil chemical sample. For this relationship, \texttt{ExperimentalUnit} is the subject and \texttt{SoilChemicalSample} is the object. Listing~\ref{lst:ontology-definitions} shows how \texttt{SoilPhysicalSample}, \texttt{bulkDensity\_g\_per\_cm\_cubed}, and \texttt{hasChemSample} are defined in the ontology using Turtle format.

\begin{figure}[h!]
    \centering
    \vspace{-0.7 cm}
    \begin{minipage}{.9\textwidth}
        \begin{lstlisting}[language=Turtle, basicstyle=\ttfamily\tiny, caption={Definitions of class SoilPhysicalSample, data property bulkDensity\_g\_per\_cm\_cubed, and object property hasChemSample.}, label={lst:ontology-definitions}]
@prefix sockg: <https://idir.uta.edu/sockg-ontology/docs/> .

###  https://idir.uta.edu/sockg-ontology/docs/SoilPhysicalSample
sockg:SoilPhysicalSample rdf:type owl:Class ;
    rdfs:comment `Represents a sample of soil collected from a specific depth range encompassing various physical properties essential for agricultural analysis and land management.' ;
    rdfs:seeAlso <https://lod.nal.usda.gov/nalt/5142> .

###  https://idir.uta.edu/sockg-ontology/docs/bulkDensitySd_g_per_cm_cubed
sockg:bulkDensitySd_g_per_cm_cubed rdf:type owl:DatatypeProperty ;
    rdfs:domain sockg:SoilPhysicalSample ;
    rdfs:range xsd:float ;
    rdfs:seeAlso <https://lod.nal.usda.gov/nalt/20349> .

###  https://idir.uta.edu/sockg-ontology/docs/hasChemSample
sockg:hasChemSample rdf:type owl:ObjectProperty ;
    rdfs:domain sockg:ExperimentalUnit ;
    rdfs:range sockg:SoilChemicalSample .

        \end{lstlisting}
        \vspace{-.9 cm}
    \end{minipage}%
 \end{figure}

The dataset's specialized nature necessitated an intricate naming convention. For clarity and consistency, class names were formatted in upper camel case (e.g., \texttt{ExperimentalUnit}), while data properties and object properties followed lower camel case (e.g., \texttt{hasAmendment}). Data properties that include units are formatted with the units separated by underscores (e.g., \texttt{aboveGroundBiomass\_kg\_per\_ha}). 
To make the ontology accessible to all users, abbreviations and acronyms that are not universally recognized and unambiguous
were avoided. For instance, acronyms such as ``FAME,'' ``PLFA'' and ``POM,'' seen in Figure~\ref{fig:ambiguous}, were expanded to their full terms \sloppy \texttt{fattyAcidMethylEsters}, \texttt{phospholipidFattyAcids} and \texttt{particulateOrganicMatter\_gC\_per\_kg}, respectively. Similarly, ambiguous terms were clarified using contextual notes (cf. Section~\ref{sec:data}). For instance, the term ``Iden Plant Material gC/kg'', also in Figure~\ref{fig:ambiguous}, was modeled as the data property \texttt{organicPlantMaterial\_gC\_per\_kg}, reflecting the description provided in the contextual note. On the other hand, the term ``DNA'', as also seen in Figure~\ref{fig:ambiguous}, is common knowledge; thus, it was simply modeled as the data property \texttt{soilDna}.
These choices demonstrate the ontology's focus on clarity and minimizing potential confusion, ensuring that it remains intuitive and precise for all users.

\vspace{1mm}
\noindent \textbf{An example of design choices in ontology development}\hspace{0.3cm} 
As described in Section~\ref{sec:data}, soil samples are used to measure various physical, chemical, and biological properties of soil within an experimental unit. These samples have attributes such as sample depth, which includes both upper and lower depth boundaries. Initially, we considered a design that defined the following classes: \texttt{ExperimentalUnit}, \texttt{SoilSample}, \texttt{SoilPhysicalMeasure}, \texttt{SoilChemicalMeasure}, and \texttt{SoilBiologicalMeasure}. Corresponding object properties included \texttt{hasSample} (linking \texttt{ExperimentalUnit} to \texttt{SoilSample}) and \texttt{hasMeasure} (linking \texttt{SoilSample} to one or more of \texttt{SoilPhysicalMeasure}, \texttt{SoilChemicalMeasure}, or \texttt{SoilBiologicalMeasure}). 
However, as mentioned earlier in Section~\ref{sec:data}, further analysis of the data and consultations with domain scientists revealed that the AgCROS dataset does not reflect, for instance, whether the \texttt{SoilPhysicalMeasure} and \texttt{SoilChemicalMeasure} recorded at the same depth and the same time from the same experimental unit were from the same real-world sample or not. To address this, we abandoned the design of the classes \texttt{SoilSample}, \texttt{SoilPhysicalMeasure}, \texttt{SoilChemicalMeasure}, and \texttt{SoilBiologicalMeasure}. Instead, we introduced distinct classes for each type of sample: \texttt{SoilPhysicalSample}, \texttt{SoilChemicalSample}, and \texttt{SoilBiologicalSample}. This approach avoids separate measure definitions by directly associating each sample type with the experimental unit through the object property \texttt{hasSample}. 

\subsection{Knowledge Graph Population} 
SOCKG is populated from the AgCROS spreadsheet through an automated two-step process. The first step involves preprocessing the spreadsheet and creating ontology mappings aligned with the data models. The second step focuses on formulating queries to construct SOCKG and storing it in graph databases (GraphDB and Neo4j in this study).

\noindent \textbf{Data preprocessing}\hspace{0.3cm} 
The AgCROS spreadsheet includes domain-specific terminology, including chemical abbreviations and shortened terms that could be misinterpreted and cause confusion. To address this issue, a mappings dictionary was created to standardize the column names by mapping them to their corresponding terms in the ontology. This helps with querying the SOCKG, as described in Section~\ref{sec:usecases}, and exploring it via the dashboard described in Section~\ref{sec:applications}.
In addition to the mappings dictionary, we assign unique IDs (UIDs) to each row in \textit{measurement} tabs. These UIDs are necessary because each data instance is inserted into the SOCKG based on a unique identifier. Using existing values from the AgCROS spreadsheet as UIDs is not feasible due to repeated values and inconsistencies across different tabs. To address this, we implemented a new column that concatenates multiple columns based on input from domain scientists. Lastly, we used the value ``NaN'' to represent missing values in the spreadsheet, distinguishing them from empty values denoted by ``None''. The distinction between empty and missing values is determined based on notes provided by domain scientists.

\vspace{1 mm}
\noindent \textbf{Data loading}\hspace{0.3cm}
The data loading process consists of converting the processed data into a structured RDF file. To begin, the ontology is parsed using RDFlib\footnote{\scriptsize\url{https://github.com/RDFLib/rdflib}}, generating an RDF graph that contains the ontology's \textit{classes}, \textit{data properties}, and \textit{object properties}. This graph provides the structure needed to build an RDF graph of the processed data. Next, entities are added by going through the ontology's classes and assigning their corresponding data properties. After the entities are added, object properties are added using a similar approach. Once the graph is built, it can be serialized into any RDF format, such as Turtle, N-Triples, or JSON-LD/XML, and then exported to a database such as Neo4j or GraphDB for analysis. 

\vspace{-0.4 cm}
\subsection{NALT Integration and Accessibility}\label{sec:constExtra}

\textbf{NALT integration}\hspace{0.3cm}
Following the design of the ontology, we prioritized integrating the terms from our ontology with those in the National Agricultural Library Thesaurus (NALT), a widely recognized controlled vocabulary in agriculture developed and maintained by USDA. This integration is crucial for achieving interoperability, standardizing terminology, and enhancing compatibility with other datasets that reference NALT. The mapping process involved mostly manual curation to verify accuracy. The primary challenges involved addressing terms in the ontology that represented broader or combined concepts, which often did not align directly with individual NALT terms, thus preventing straightforward one-to-one mappings.

Approximately 61\% of SOCKG’s classes and properties have been aligned with NALT. For integration, we utilize the built-in \texttt{rdfs:seeAlso} annotation property in Protégé to provide supplementary links to related or equivalent concepts. For instance, the NALT term corresponding to the class \texttt{Treatment} is ``experimental treatments'', with the URI https://lod.nal.usda.gov/nalt/6148134.
For our future work, we plan to utilize additional OWL constructs, such as \texttt{owl:equivalentClass} to establish precise equivalences between SOCKG ontology classes and NALT terms and \texttt{owl:equivalentProperty} to link equivalent properties between the ontology and NALT. This ongoing integration effort will lead to a strong foundation for improving data consistency across agricultural research platforms. 

\vspace{1mm} 
\noindent \textbf{Resolvable URIs}\hspace{0.3cm}
To improve clarity, accessibility and interoperability, all classes, object properties and data properties in SOCKG ontology are made resolvable through URIs. These URIs enable users to retrieve detailed information about each entity and facilitate integration with external datasets. Designed for human readability, the URI structure follows the format https://idir.uta.edu/sockg-ontology/docs/{conceptName}, where {conceptName} can represent the name of a class, an object property, or a data property. These URIs are hosted on a managed publicly accessible server to ensure reliability and long-term availability. Users can access the resolvable pages via a browser, where they can view metadata such as descriptions, equivalent NALT terms, related object properties of each class, and the domain and range of each data property. 
Currently, only classes and properties have resolvable URIs, but work is in progress to extend this functionality to individual data instances. 

\vspace{-0.3 cm}
\section{Query Endpoints, Dashboard, and Data Cube}
\label{sec:applications}
\subsection{Query Endpoint}
SOCKG is stored in two widely used graph databases---Neo4j and GraphDB---both with dedicated browsers for user access. GraphDB, a semantic graph database, supports RDF data and enables users to query using SPARQL, making it suitable for applications requiring complex semantic search and reasoning capabilities. Neo4j, on the other hand, is a property graph database optimized for highly connected data, allowing users to query through the Cypher query language, which is well-suited for traversing relationships in large networks. 

The inherent complexity of the semantic web and the steep learning curve of SPARQL can limit access to semantic data, especially for users unfamiliar with query languages. To make SOCKG more accessible, we developed an interactive dashboard that simplifies data exploration and analysis.

\vspace{-0.3cm}
\subsection{Dashboard}
The dashboard is built using the Streamlit framework\footnote{\scriptsize\url{https://streamlit.io/}}, which provides a visual interface that abstracts the underlying complexity and organizes the knowledge graph into multiple pages, each dedicated to a specific data region. The following is an overview of the design and functionality of the dashboard's pages.

\begin{figure}[h!]
    \centering
    \begin{minipage}{0.48\textwidth}
        \centering
        \vspace{-0.4cm}
        \includegraphics[width=\linewidth]{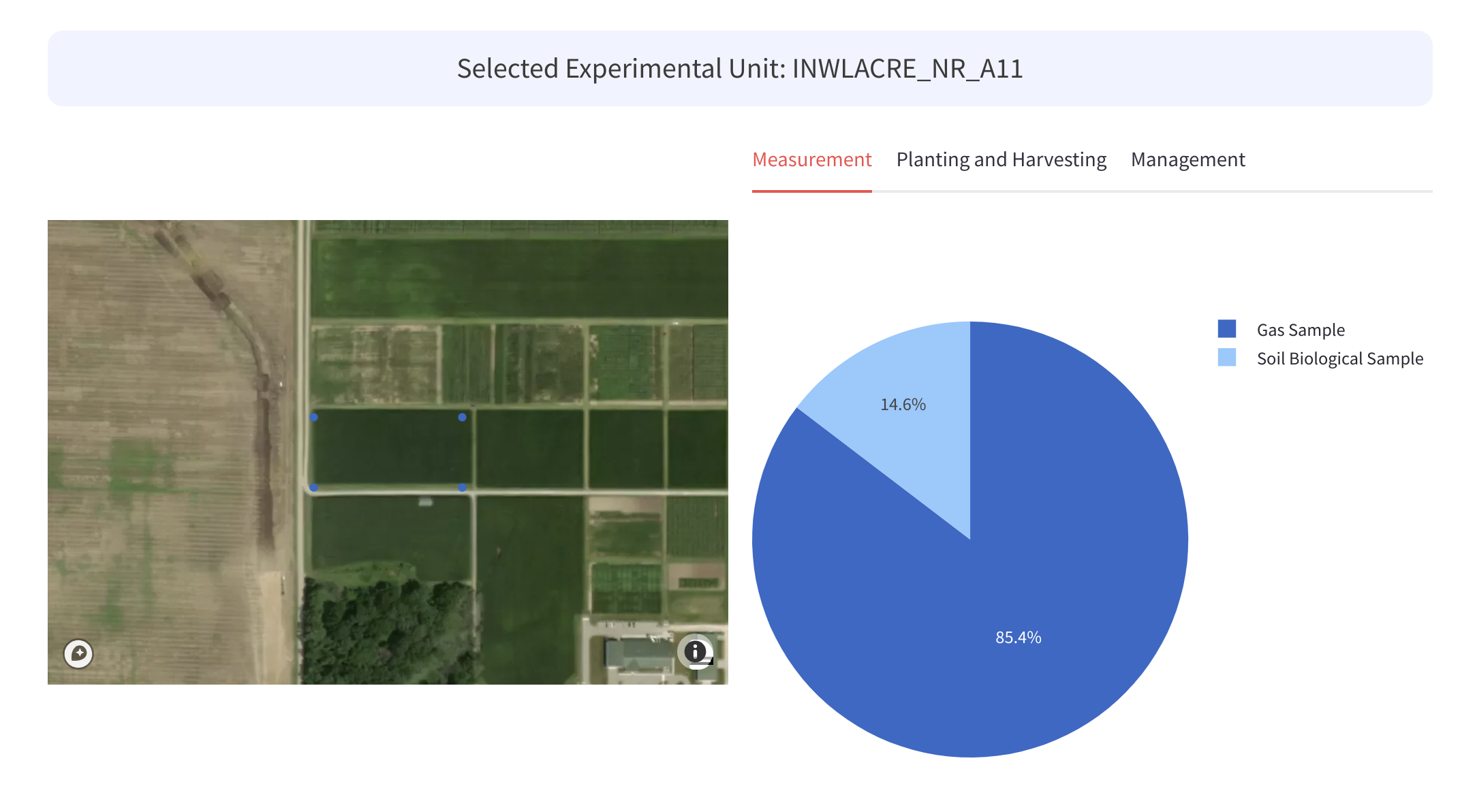}
        \vspace{-0.4cm}
        \caption{Satellite image and pie chart displaying soil sample count of experimental unit \texttt{INWLACRE\_NR\_A11}.}
        \label{fig:dashboard-exp-geo}
        \vspace{-0.4cm}
    \end{minipage}%
    \hfill
    \begin{minipage}{0.48\textwidth}
        \centering
        \vspace{-0.4cm}
        \includegraphics[width=\linewidth]{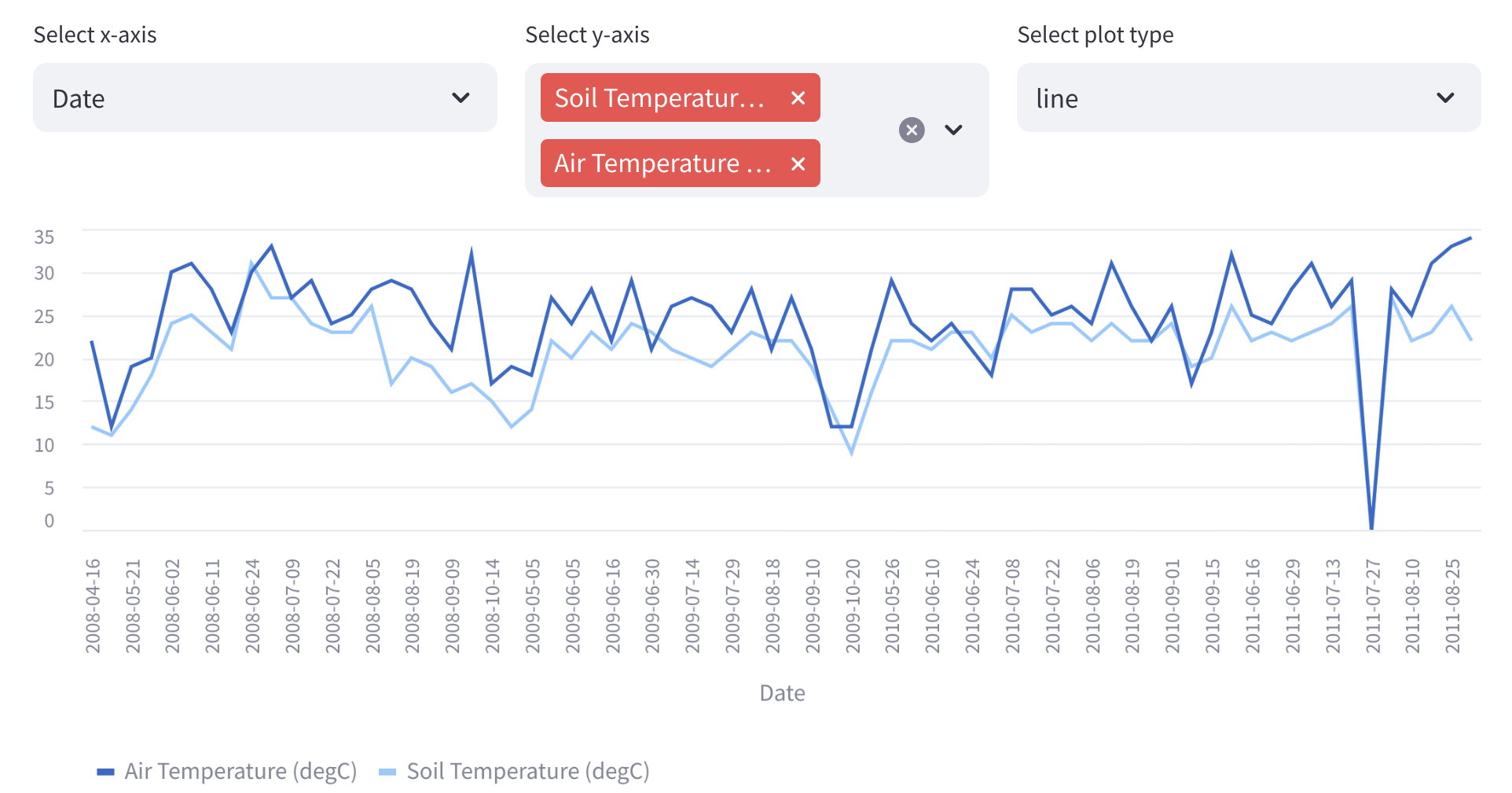}
        \vspace{-0.4cm}
        \caption{A 2D line chart with the x-axis representing the dates and the y-axis representing soil temperature and air temperature (in degrees) collected from the experimental unit \texttt{INWLACRE\_NR\_A11}.}
        \label{fig:dashboard-exp-plot}
        \vspace{-0.4cm}
    \end{minipage}
\end{figure}

\noindent \textbf{Experimental unit exploration}\hspace{0.3cm}
This page visualizes SOCKG data across the U.S. using a heat map, where darker colors indicate states with more experimental units. Users can apply filters (State, County, Site, Field) to refine their search. Clicking on an experimental unit shows its location on the map (if available) and a pie chart of soil sample counts (Figure~\ref{fig:dashboard-exp-geo}). Users can also view soil samples in a table and plot data from selected columns using customizable chart types (Figure~\ref{fig:dashboard-exp-plot}).

\begin{figure}[h!]
    \centering
    \begin{minipage}{0.53\textwidth}
        \centering
        \includegraphics[width=\linewidth]{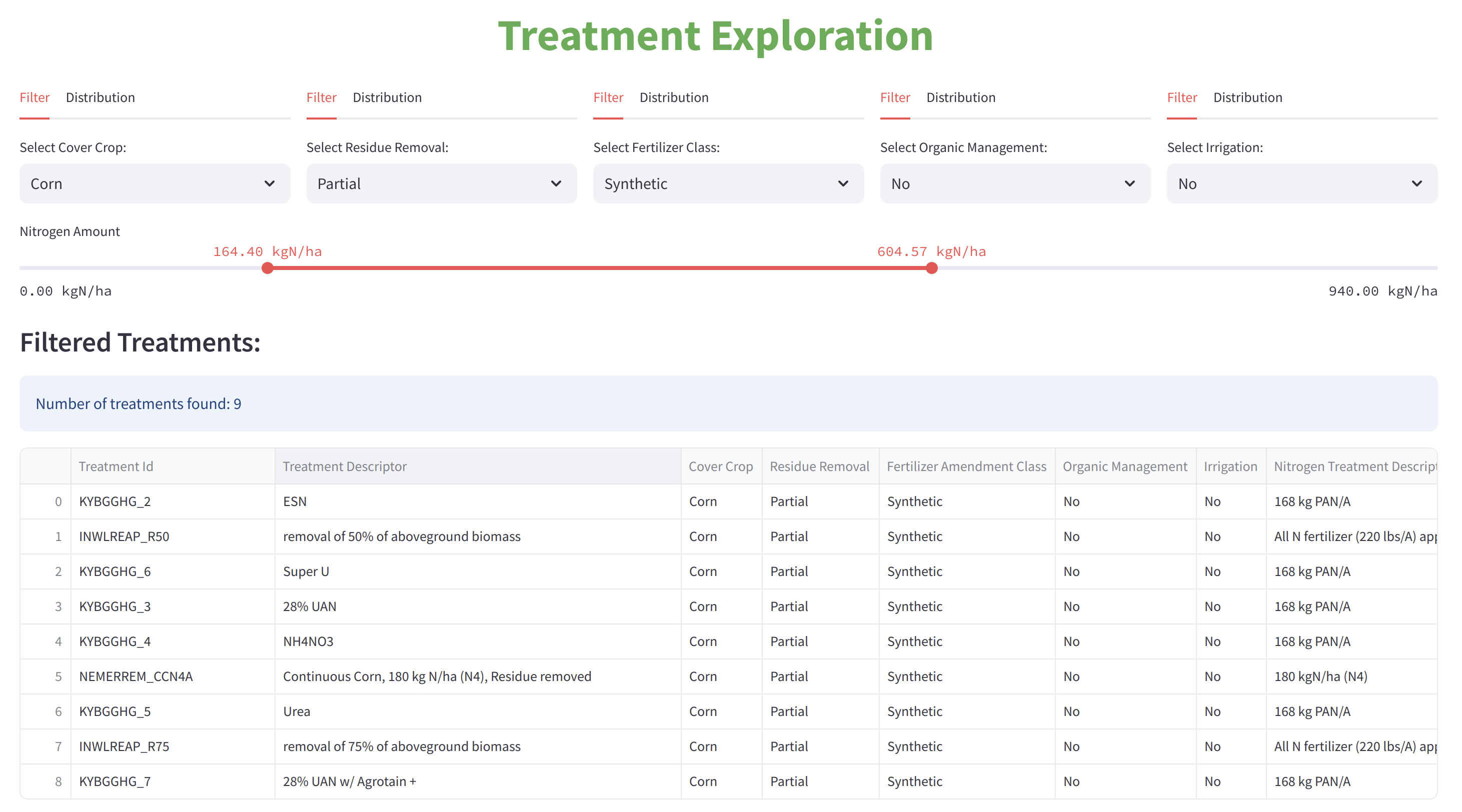}
        \vspace{-0.4cm}
        \caption{Filtered treatments meeting criteria such as crop type (Corn), residue removal (Partial), and fertilizer class (Synthetic).}
        \label{fig:dashboard-treatment}
        \vspace{-0.6cm}
    \end{minipage}%
    \hfill
    \begin{minipage}{0.45\textwidth}
        \centering
        \includegraphics[width=\linewidth]{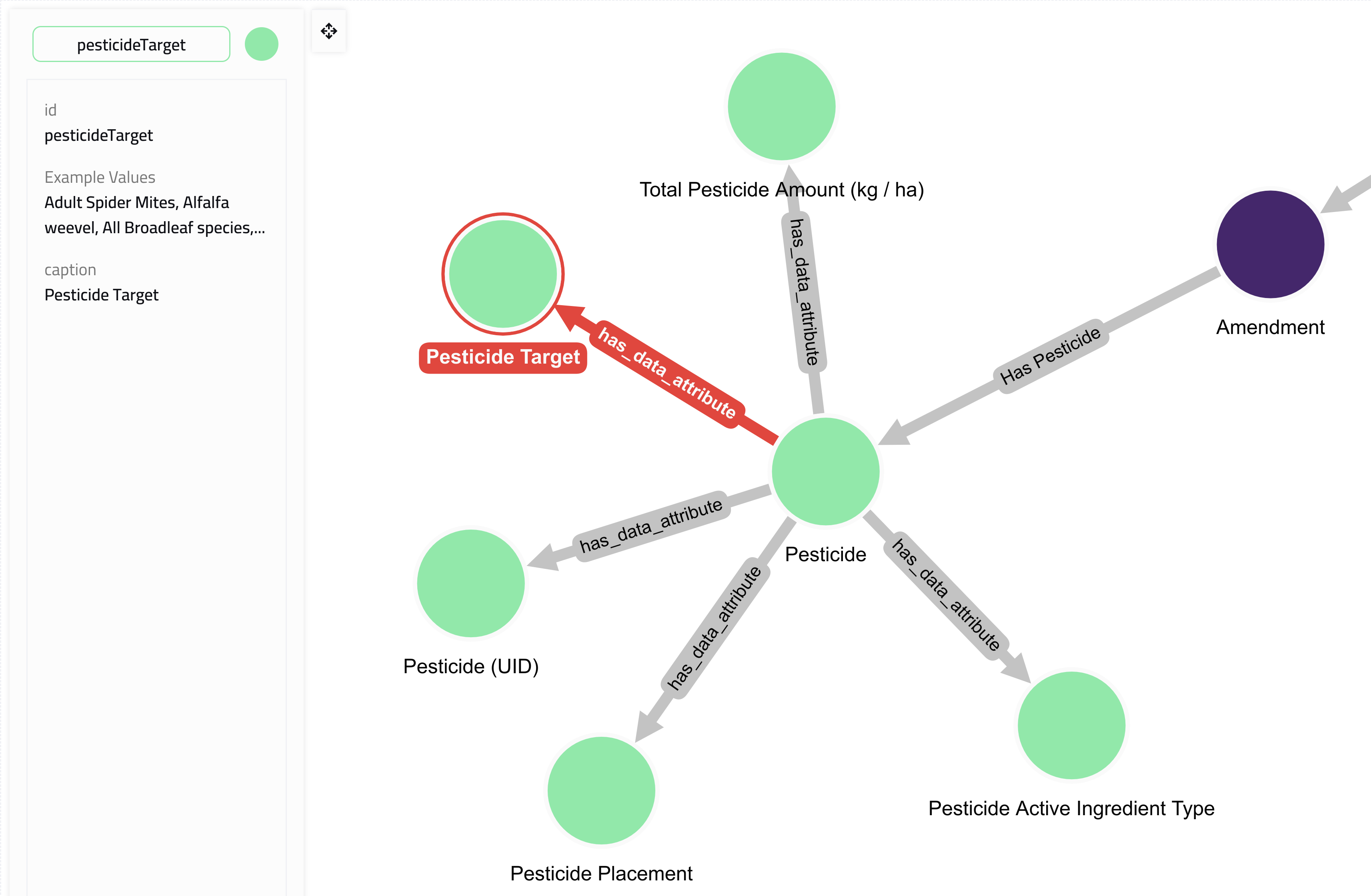}
        \vspace{-0.4cm}
        \caption{Expanded Pesticide class with its corresponding data properties (e.g., pesticideTarget) and example property values (e.g., ``Adult Spider Mites,'' ``Alfalfa Weevel,'' and ``All Broadleaf Species'').}
        \label{fig:dashboard-onto-expanded}
        \vspace{-0.6cm}
    \end{minipage}
\end{figure}

\noindent \textbf{Treatment exploration}\hspace{0.3cm} This page helps users explore treatments using filters such as crop type and fertilizer type (organic, synthetic, etc.), as shown in Figure~\ref{fig:dashboard-treatment}. The search interface is designed according to the principles of a faceted interface~\cite{faceted-interface}---whenever a value is chosen in a drop-down box, the available options in other drop-down boxes will be updated accordingly.  
Once a treatment is selected, all experimental units associated with that treatment are displayed, and the user can jump from current page to an experimental unit page by clicking.


\noindent \textbf{Ontology exploration}\hspace{0.3cm} This page (Figure~\ref{fig:dashboard-onto}) offers users a visual overview of the SOCKG ontology. A user can click on any node to view the total number of instances in that class (e.g., 3,809 experimental units). By double clicking a class, the user can see its data properties, and by clicking an data property, they can view the corresponding sample values and data types (as shown in Figure~\ref{fig:dashboard-onto-expanded}). The ontology explorer was also beneficial in validating the ontology in its development process, helping identify missing relationships and verifying the consistency of class relationships.

\subsection{Data Cube}\label{sec:datacube}
\noindent \textbf{Motivation}\hspace{0.3cm}
While representing soil carbon data as a knowledge graph is essential for integrating it with the broader semantic web, addressing analytical questions from end users, such as \textit{what is the influence of certain agricultural practices on soil health?}, is not inherently straightforward using data in RDF format due to several reasons. \textit{First}, RDF stores data in triples, and queries are similar to graph traversal, meaning that execution time often correlates with the number of possible paths (both incoming and outgoing) from the current nodes. Unfortunately, SOCKG has a high degree of centralization on the Experimental Unit node (see Figure \ref{fig:dashboard-onto}). Analytical questions often require examining multiple experimental units and their measurements. This traversal is costly, as many experimental units have lifespans of up to 15 years and take multiple measurement samples at different depths daily, resulting in an enormous number of outgoing degrees. In our experiments, this traversal process took an average of around 15 seconds for such an analytical question. \textit{Second}, RDF can be queried using standard query languages such as SPARQL. In practice, we found a simple question such as \textit{What is the average SOC stock (kgC/ha) for 0-30 cm?} would require around 45 lines of SPARQL code in its most compact form. The primary challenges stem from the need to derive analytical data through the aggregation of raw database records, as well as the reliance on implicit knowledge for calculations. For instance, computing SOC stock (kgC/ha) requires applying formulas involving organic carbon content, bulk density, soil depth, and unit conversions (e.g., 1 ha = 10,000 m\(^2\)). Translating these calculations into SPARQL code involves multiple steps and intermediate results.

\noindent Given these challenges, we identified two key objectives: simplifying SPARQL query writing and minimizing response time for analytical questions. To address these, we utilize multidimensional data cubes~\cite{gray1997data} to model soil carbon questions, incorporating factors such as weather, soil management practices, and soil type, each represented as a dimension in the data cube.

\noindent \textbf{Implementation}\hspace{0.3cm}
While tools like Apache Kylin\footnote{\scriptsize\url{https://kylin.apache.org/}} support both data cube modeling and precomputation, they are not a good fit for our use case. Kylin requires data to be structured in relational tables and, by default, precomputes the entire data cube. This approach is well-suited for industrial applications but is excessive for our needs. Fully precomputing the data cube would demand significant storage space, as the required storage grows exponentially with the number of dimensions. Therefore, our design seeks a balance between full precomputation, which is storage-intensive, and on-the-fly query processing, which requires noticeable CPU computation time.
Therefore, we leverage the concept of the data cube to systematically structure such analytical questions. 
For example, a data cube for SOC treats soil organic carbon as the fact, with influencing factors as dimensions. The same approach applies to harvest yield or greenhouse gas emissions (with slightly adjusted dimensional tables). Once the data cube model is created, we can easily identify various analytical question instances at different granularity levels by examining the dimension tables. This allows us to precompute results for common queries and store them for later retrieval. The entire process is described below.

\textbf{Dimension value selection}\hspace{0.3cm} For each dimension table (e.g., time, location, management), we define a set of relevant values at a reasonable granularity based on domain scientists' input. For example, in the management dimension, the user might be interested in scenarios with no tillage or with disk tillage. This step involves the most manual work, as it is challenging to determine the optimal granularity and it is impossible to account for all possible values, especially for continuous dimensions such as time. However, it remains a trade-off between performance and completeness, as discussed earlier.

\textbf{Combination generation}\hspace{0.3cm} Using the values identified for each dimension, we write a script to iterate over all possible combinations. This generates a list of all potential combinations of dimension values, with the option to include an empty value for each set. For example, a combination might be \textit{no tillage}, \textit{synthetic fertilizer}, \textit{use irrigation}, corresponding to the question: \textit{``What is the average SOC stock on experimental units with no tillage but irrigation?''} In practice, the combinations are more complex, as we have 9 dimension tables, each with about 3 attributes.

\textbf{Pre-computation and result storage}\hspace{0.3cm} Using the generated list of combinations, we insert them into a pre-defined SPARQL template to fetch the corresponding results. We then store all combination-result pairs in a NoSQL database, allowing for constant-time lookups during query time through our custom API.

\begin{figure*}[h!]
    \centering
    \begin{minipage}{0.49\textwidth}
        \centering
        \vspace{-0.2cm}
        \includegraphics[width=\linewidth]{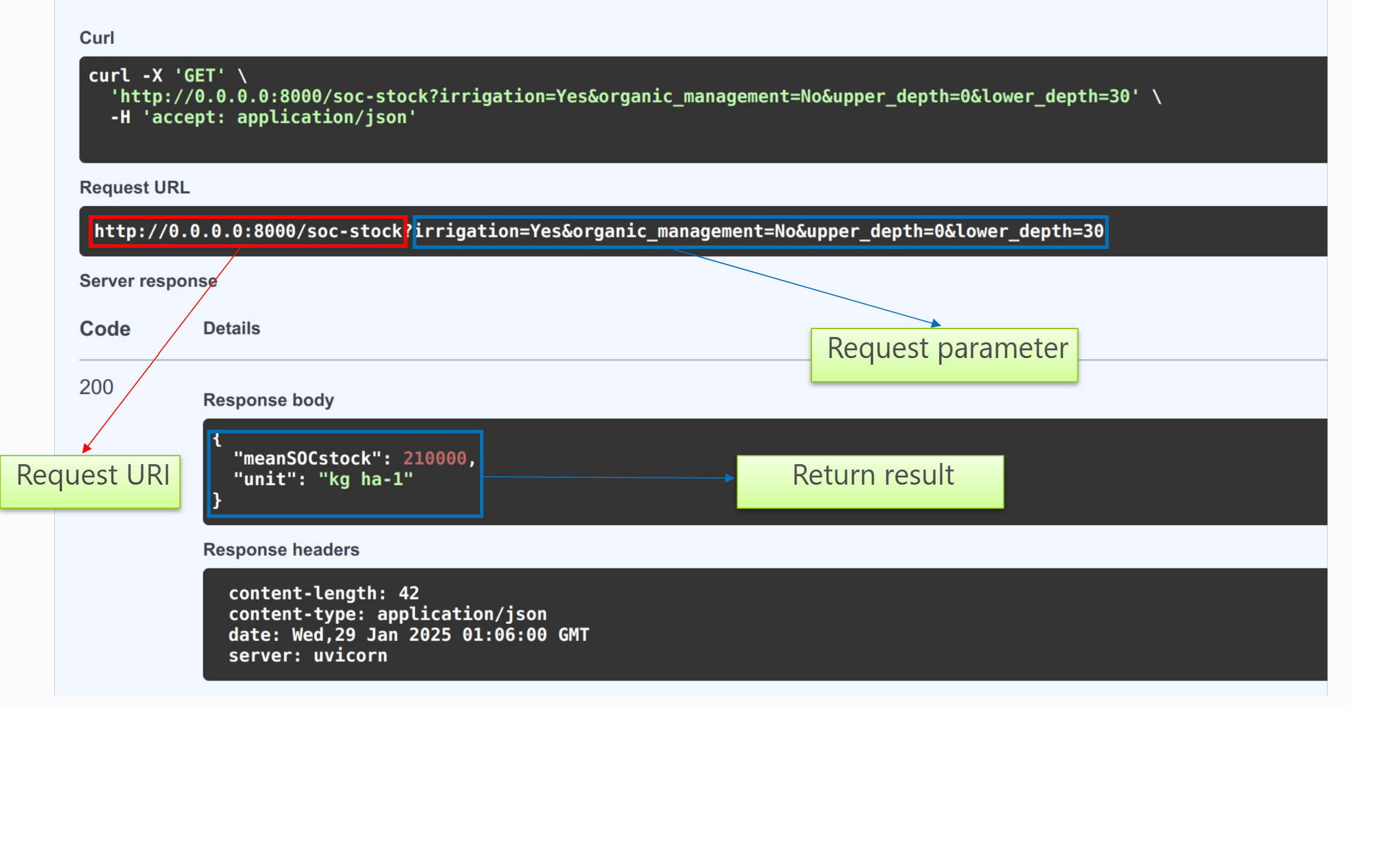}
        \vspace{-1.2cm}
        \caption{Example usage of the data cube API to retrieve the average SOC stock for all experimental units where irrigation is applied, measured at a depth of 0--30 cm.}
        \label{fig:fast-api}
    \end{minipage}\hfill
    \begin{minipage}{0.49\textwidth}
        \centering
        \includegraphics[width=\linewidth]{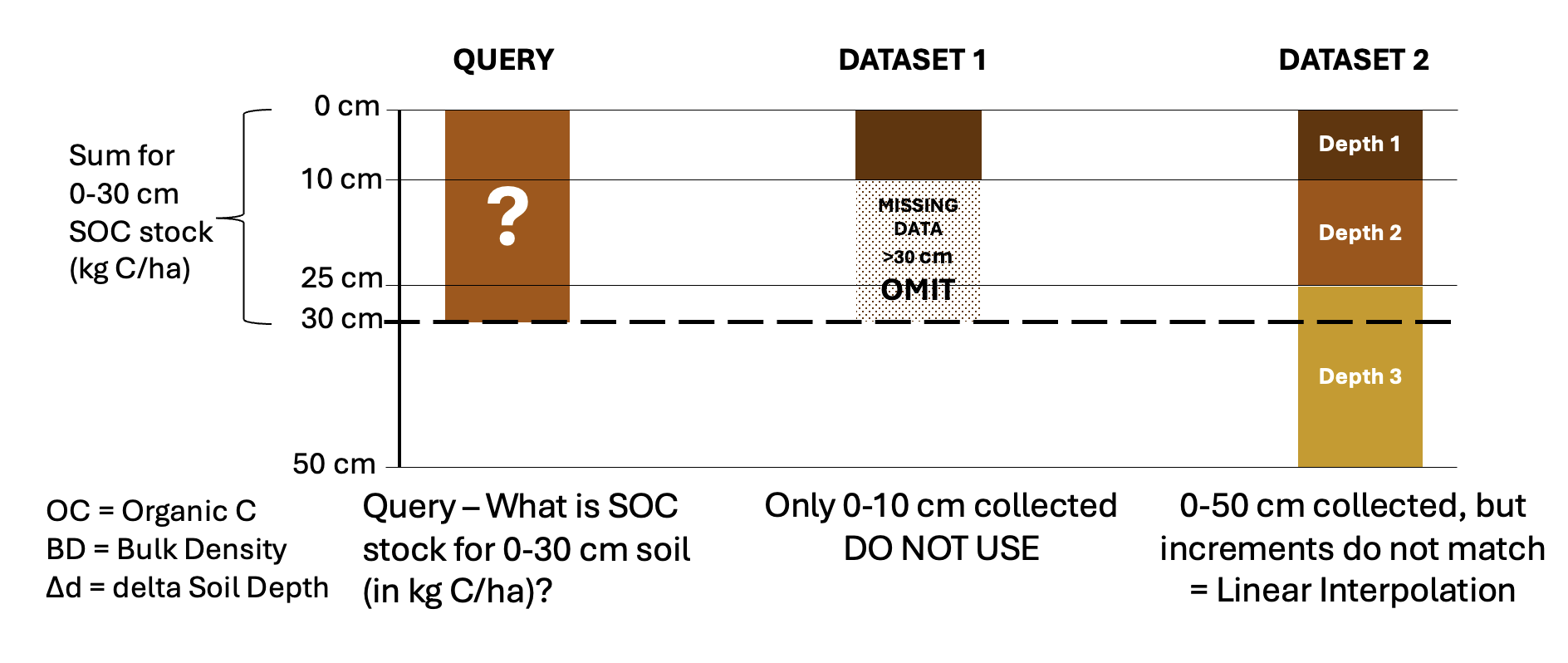}
        \vspace{-0.4cm}
        \caption{An example of linear interpolation where the equation \(\sum_{i} (OC_i \cdot BD_i \cdot \Delta SD \cdot 100)\) applies to layers between 0--30 cm.}
        \label{fig:linear_interpolation}
    \end{minipage}
    \vspace{-1.1cm}
\end{figure*}

Using the above pipeline, we implemented a test version with a simplified data cube design (i.e., using soil carbon stock as the fact) via FAST API. For this version, we limited the data included, and storing pre-computed data in a database showed minimal benefit, so we simplified the process. Currently, we provide a REST API that can be used by both end-users and dashboard developers, with dimensional values flattened into GET request parameters; an example is shown in Figure \ref{fig:fast-api}. In the future, we aim to refine the data cube design and cache results in a NoSQL database.
\vspace{-0.5 cm}

\section{Use Cases and Discussion}
\label{sec:usecases}
\vspace{-0.2 cm}
This section highlights how SOCKG can support scientists in answering their research questions through an illustrative query provided by the USDA soil scientists. This query was created to address vital SOC stock questions that allow for deeper understanding of how to increase crop productivity and decrease carbon within the atmosphere. The query seeks to determine \textit{the average SOC stock (kgC/ha) for every treatment ID between 0-30 cm} where \textit{SOC stock (kgC/ha) = Organic C percentage (gC/kg~soil) * Bulk Density(g~soil/cm$^{3}$) *  \(\Delta\)Soil Depth(cm) * 100}. Due to data variability, the query required handling several edge cases which, if unaddressed, could compromise the validity of the results. The most important case is the variability in parameters in data collections, such as depth, which are not always consistent. For example, some samples extend only 10 cm into the soil, or employ depth partitions that differ from other intervals such as 0–10 cm, 10–15 cm, and 15–30 cm. Thus, we must consider a few scenarios that require further calculation or filtering. Samples that do not reach the user-defined lower depth, 30 cm in this case, are filtered out of the results. Furthermore, if a sample's layers do not align with the 30 cm mark, linear interpolation is applied to the layer containing 30 cm to calculate the SOC stock (see Figure~\ref{fig:linear_interpolation} for details). According to the USDA soil scientists, obtaining valid results requires adherence to following standardized calculation procedure. First, the SOC stock for each layer within the 0–30 cm range must be calculated and summed across all layers in the same sample. The summed SOC stock values are then averaged across similar treatment IDs, fields, or other relevant groupings for analysis. Given the aforementioned query, these values will be averaged based on treatment IDs. 

Upon completing this query, further analyses can be performed to extract meaningful insights---for example, ranking treatment IDs or methods based on their impact on increasing SOC stock in the soil. Through the integration and standardization of diverse data sources, SOCKG makes it possible to generate these queries and others of a similar nature, providing scientists and decision makers with accurate data and insights on soil carbon stocks, fluxes, and dynamics. SOCKG facilitates large-scale research to enhance carbon sequestration strategies and, through advanced querying and potential machine learning integration, improves predictions of soil carbon stocks while reducing uncertainties.
\vspace{-0.5 cm}
\section{Conclusion and Future Work}
\label{sec:conclusion}
\vspace{-0.3 cm}
We developed SOCKG, a semantic infrastructure designed to integrate and analyze diverse soil organic carbon data. By creating a robust ontological model and aligning it with the NALT, SOCKG ensures interoperability and consistent terminology across datasets. Implemented in GraphDB and Neo4j and complemented by an intuitive dashboard and a data cube, it enables advanced analyses, such as cross-field comparisons of soil carbon dynamics. SOCKG offers a resource for studying carbon sequestration and informing sustainable agricultural practices, with potential relevance for researchers, land managers, and policymakers. Future work includes expanding the dataset, enhancing analytical capabilities, and deepening integration with global agricultural research networks. 
\vspace{-0.4 cm}
\subsubsection{\ackname}
This material is based upon work supported by the National Science Foundation under Grants 
TIP-2333834.


\end{document}